\documentclass[%
reprint,superscriptaddress,
amsmath,amssymb,
aps,
prb,
showpacs,
]{revtex4-1}
\usepackage{xcolor}
\usepackage{amssymb}
\usepackage{graphicx}
\usepackage{dcolumn}
\usepackage{bm}
\usepackage{hyperref}

\begin{document}

\title{Efficient thermoelectric materials using nonmagnetic double perovskites
       with $d^0$/$d^6$ band filling}

\author{Pablo Villar Arribi}
\affiliation{Departamento de F\'{\i}sica Aplicada, 
              Universidade de Santiago de Compostela, 
              E-15782 Santiago de Compostela, Spain}
\affiliation{Instituto de Investigaci\'ons Tecnol\'oxicas, 
              Universidade de Santiago de Compostela,  
              E-15782 Santiago de Compostela, Spain}
\affiliation{European Synchrotron Radiation Facility,
             71 Avenue des Martyrs, F-38000 Grenoble, France}

\author{Pablo Garc\'{\i}a-Fern\'andez}
\affiliation{Departamento de Ciencias de la Tierra y F\'{\i}sica
              de la Materia Condensada, Universidad de Cantabria, 
              Cantabria Campus Internacional,
              Avenida de los Castros s/n, E-39005 Santander, Spain}
							
\author{Javier Junquera}
\affiliation{Departamento de Ciencias de la Tierra y F\'{\i}sica
              de la Materia Condensada, Universidad de Cantabria, 
              Cantabria Campus Internacional,
              Avenida de los Castros s/n, E-39005 Santander, Spain}

\author{Victor Pardo}
\email{victor.pardo@usc.es}	
\affiliation{Departamento de F\'{\i}sica Aplicada, 
              Universidade de Santiago de Compostela, 
              E-15782 Santiago de Compostela, Spain}
\affiliation{Instituto de Investigaci\'ons Tecnol\'oxicas, 
              Universidade de Santiago de Compostela,  
              E-15782 Santiago de Compostela, Spain}
						
\date{\today}

\begin{abstract}
 Efficient thermoelectric materials should present large Seebeck coefficient, 
 high electrical conductivity and low thermal conductivity. 
 An enhanced Seebeck coefficient can be obtained from materials where the
 Fermi level can be aligned with a large and narrow peak of the density of states, particularly when a substantial band valley degeneracy occurs.
 A high electrical conductivity comes as a consequence of large 
 conductive hopping paths between the atoms of the material.
 Both physical quantities can be decoupled and optimized  
 independently if their origins can be ascribed to different sets of bands.
 Based on these assumptions,
 double perovskites A$_2$BB'O$_6$ with $d^0/d^6$ filling for the 
 B and B' metal cations, respectively, have been considered.
 They provide a desirable band structure with degenerate B-$t_{2g}$ / B'-$e_g$ 
 bands above the Fermi level together with a low thermal conductivity.
 We have carried out first-principles simulations for various of 
 these nonmagnetic double perovskites and showed that all of
 them present a large Seebeck coefficient (consequence of the localized
 and empty $t_{2g}$ states of the B-cation),
 and large electrical conductivity due to the more spread unoccupied 
 $e_{g}$ band of the B' cation.
 We have seen that if they can be optimally doped, they could show a figure of
 merit comparable or even higher than the best $n$-type thermoelectric oxides,
 such as SrTiO$_{3}$.
 Different mechanisms to tune the band structure 
 and enhance the thermoelectric figure of merit are explored, including
 epitaxial strain, hydrostatic pressure, chemical pressure, and external
 doping. A fully relaxed structure has also been studied, showing that a realistic calculation
 is necessary to make accurate predictions, but also proving that the main trends shown throughout the paper remain unchanged.
\end{abstract}



\maketitle

\section{Introduction}
\label{sec:intro}
 The quest for new materials with outstanding thermoelectric (TE) 
 properties is one of the cutting-edge
 research topics nowadays because of its multiple applications: 
 from Peltier cooling to the conversion of 
 waste heat into electricity in order to, for instance, reduce fuel 
 consumption in vehicles.
 The TE efficiency of these materials can be characterized by a 
 dimensionless TE figure of merit $zT=\sigma S^2 T / \kappa$,
 where $\sigma$ is the electrical conductivity, $S$ is the
 Seebeck coefficient, $T$ is the absolute temperature, 
 and $\kappa=\kappa_e+\kappa_l$ is the thermal conductivity
 which contains both the electron, $\kappa_e$, and lattice, $\kappa_l$, 
 contributions. 

 Improving $zT$ can be achieved combining different strategies.
 A first choice consists in reducing the total thermal conductivity, 
 which often implies reducing the lattice part.
 Heat carried by lattice vibrations (phonons in the quantized form) 
 is particularly detrimental to TE performance\cite{Singh-08} since the 
 heat backflow leads to a reduction in the temperature gradient 
 required for the TE module to operate.
 The idea of reducing thermal conductivity while retaining reasonable values for both the electrical conductivity and the Seebeck coefficient
 lies at the basis of the 
 ``phonon-glass, electron-crystal'' concept developed by Slack more than
 twenty years ago.~\cite{Slack-95,Beekman-15}
 This is often carried out by
 the use of materials with very
 complex unit cells~\cite{snyder2008,Nolas} 
 or cage-like compounds in which heavy atoms are enclosed.
 The rattling of these heavy atoms strongly scatters phonons,
 specially those with a longer wavelength, 
 that are the most involved in the heat transport,
 with the corresponding reduction in lattice thermal conductivity.
 Examples of structures that can lead to excellent TE materials, 
 some of these based on these concepts are
 skutterudites,~\cite{Sales-96,Singh-97,Keppens-98} 
 clathrates,~\cite{Saramat-06} chalcogenides,\cite{chalcs} 
 Zintl-phases,~\cite{zintl} or 
 half-Heusler alloys.~\cite{Shen-01,Culp-08}

 A second way to improve the TE figure of merit 
 is to maximize the power factor, defined as the product $\sigma S^{2}$,
 by varying the carrier concentration
 with different doping levels.~\cite{heavydopping1,heavydopping2,heavydopping3}
 This approach has its limitations, since the Seebeck coefficient, electrical
 conductivity, and electronic thermal conductivity are strongly interconnected.
 Typically, $S$ is small in metals, leading to small power factors.
 On the other hand, $S$ is large for semiconductors and insulators
 but $\sigma$ is lower for them, so the power factor $\sigma S^{2}$ 
 is again very small.~\cite{Bruce} 
 For a given material, the carrier concentration can be optimized to achieve
 a good compromise among these quantities in order to enhance the value of the power factor.
 The values that maximize the power factor are usually of the order of 
 10$^{18}$-10$^{19}$ cm$^{-3}$, typical in 
 extrinsic doped semiconductors and semimetals.
 However, the interdependency of the Seebeck coefficient and the electrical
 conductivity hampers the applicability of this optimization method.
 A more desirable strategy 
 would require the decoupling of these two quantities. 

 Since the Seebeck coefficient is directly proportional to the
 energy derivative of the electronic conductivity at the Fermi energy, 
 $E_{\rm F}$, 
 itself related with the density of states (DOS),~\cite{Ashcroft} then    
 the presence of narrow and sharp features of the DOS around $E_{\rm F}$ 
 can increase the thermopower
 [see Eq. (2) of Ref.~\onlinecite{Heremans-08}].
 The idea of tuning the band structure of materials to enhance the 
 thermoelectric figure of merit was proposed by Hicks and
 Dresselhaus in two milestone works two decades
 ago.~\cite{Hicks-93.1,Hicks-93.2} Working within a free-electron model,
 these authors showed how a reduction in the dimensionality 
 of the TE materials could
 dramatically improve their performance because 
 (i) the electron confinement in low-dimensional systems modifies 
 the shape of the DOS, that displays a 
 staircase-shaped energy dependence with each step being associated
 with one of the energy states (in 2D), 
 divergence near the bottom of each of the one-dimensional subbands (in 1D) 
 or isolated peaks located at the energy states (in 0D), 
 with large derivatives at the previous particular energy positions, and
 (ii) the boundaries of the low-dimensional system would increase
 the scattering of phonons with wavelengths comparable or larger than 
 its spatial dimensions, leading to a considerable reduction of the thermal conductivity.

 Another approach to induce peaks in the DOS is the use of
 materials with strongly correlated electrons. 
 From this perspective, oxides are a very attractive option.~\cite{Hebert-15} 
 Oxygen can be combined with many different cations in 
 relatively simple structures that present very different structural 
 phase-transition sequences involving polar and non-polar distortions; 
 equilibrium phases of insulating, semi-conducting, metallic and 
 even superconducting character; 
 and piezoelectric, ferroelectric, ferromagnetic, multiferroic or 
 TE properties, making them ideal candidates
 for multifunctional devices.
 Also, they are stable under ambient conditions and they can be 
 processed on the nanoscale 
 in the form of thin films, nanoparticles, or even nanowires
 using nowadays standard techniques.
 The study of TE properties on oxides was boosted after 
 the discovery of a large thermopower in the metallic layered
 Na$_{x}$CoO$_{2}$ cobaltates, with $zT$ values up to 1.0 at 
 800 K.~\cite{thermoCo1,thermoCo2,thermoCo3,thermoCo4}
 This family of materials are composed by triangular planes of CoO$_{2}$, where the Co atoms sit inside trigonally distorted edge-sharing octahedra, separated by layers filled with Na$^{+}$. 
 Their electronic structure combines two sets of electrons produced by the 
 trigonal splitting of the t$_{2g}$ manifold: 
 delocalized $e_g^{\pi}$ and more localized $a_{1g}$ electrons.\cite{naxcoo2_es}
 The former provide a large conductivity, and the latter help increasing the 
 Seebeck coefficient.
 This, together with their relatively small thermal conductivity, 
 places layered cobaltates as the highest $zT$ recorded
 for oxides. 
 Other oxides that show large power factors are, e.g. $n$-type doped 
 SrTiO$_3$ (nominal 3$d^0$ electronic configuration) or, more recently CrN 
 (not an oxide, but very close in terms
 of electronic structure~\cite{botana2012electronic} with its nominal 
 3$d^3$ electronic configuration with a full majority-spin $t_{2g}$ 
 band in each site and its naturally occurring electron doping).
 Both of them have been found to increase its $zT$ dramatically when
 nanostructured (see Ref.~\onlinecite{ohta2007giant} for SrTiO$_{3}$
 and Ref.~\onlinecite{quintela2015epitaxial} for CrN).
 These two systems have in common the occurrence of empty $d$-bands just above 
 the Fermi energy that yield a set of localized levels that are 
 sensitive to electron doping.
 Very similar band fillings will be explored in the present work.

 \begin{figure}[!ht]
    \begin{center}
       \includegraphics[width=0.90\columnwidth]{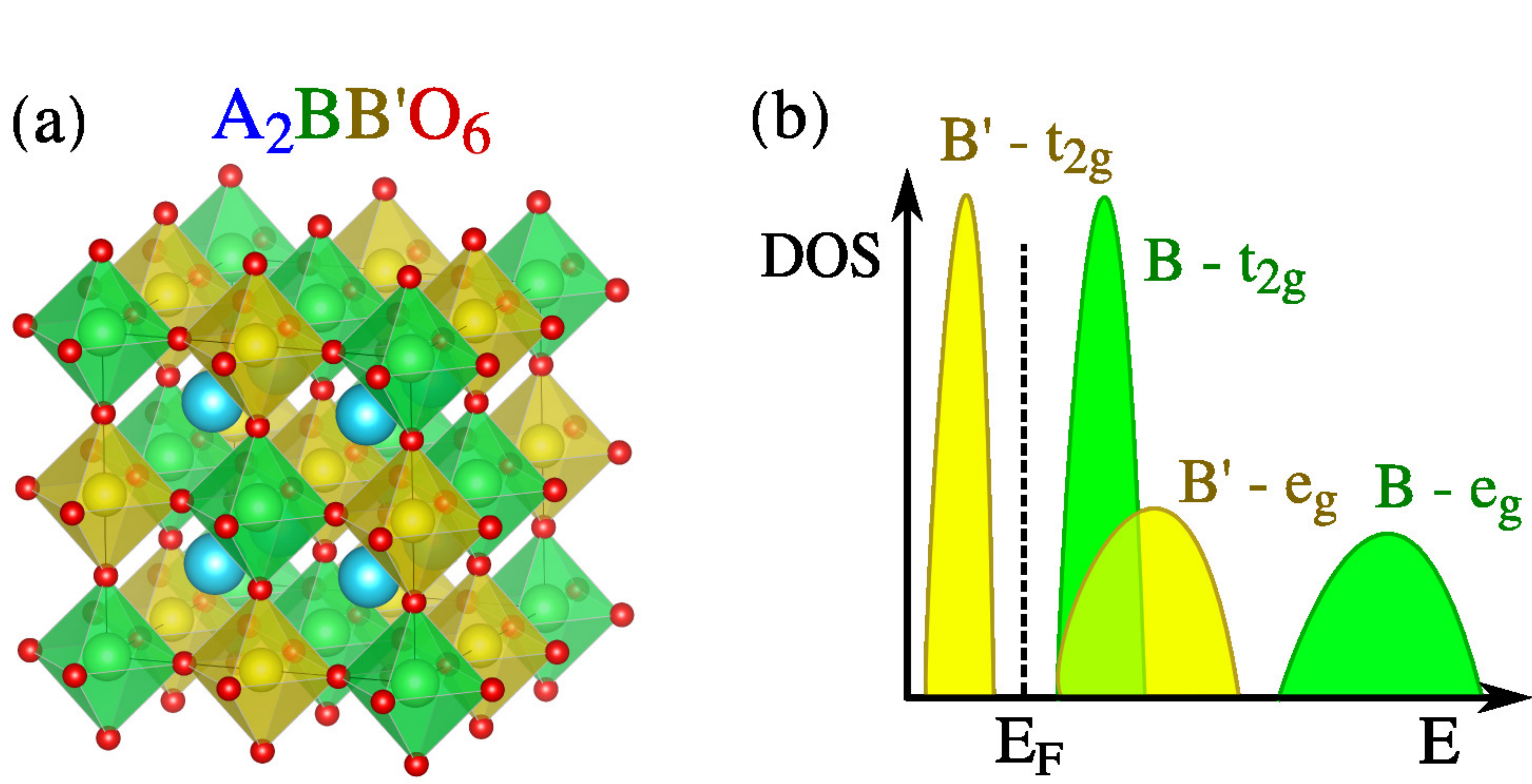}
       \caption{ (Color online) (a) Schematic representation of the 
                 A${_2}$BB'O$_{6}$ double-perovskite structure studied in this work, 
                 where A denotes a di- or trivalent cation, while B and B' are
                 transition metal atoms. 
                 O atoms are shown in red, A cation in blue, and the 
                 B and B' cations are represented by green (dark) and 
                 yellow (clear) spheres,
                 with the O octahedra colored accordingly.
                 (b) Scheme of the expected electronic structure configuration,
                 represented by the DOS,  
                 for $d^0/d^6$ band filling of the B (empty $d$ shells,
                 shown in green)
                 and B' (full $t_{2g}$ manifold displayed in yellow) cations, 
                 respectively.}
       \label{fig:perovskite}
    \end{center}
 \end{figure} 

 Trying to put together all these evidences for oxides and TE materials 
 in general,
 we have 
 attacked the problem from a different facet:
 instead of playing with dimensionality,
 we exploit the very flexible chemical composition that can be
 found in oxides.
 We focus on double perovskites of 
 generic chemical formula A$_2$BB'O$_6$ 
 [see a picture of the well-known structure in Fig. \ref{fig:perovskite}(a)],
 where A is a di- or trivalent cation, while B and B' are different transition 
 metals that are combined
 in such a way that they have a $d^0/d^6$ filling. 
 The expected DOS is schematically represented in the cartoon of 
 Fig.~\ref{fig:perovskite}(b).
 The B cation has a completely empty $d$-shell, with a localized $t_{2g}$
 band just above the Fermi level. 
 When the system is electron-doped, this band should yield a large thermopower.
 Assuming that the B' cation is in a low-spin state, it presents a full 
 $t_{2g}$ shell.
 Due to the presence of a broad empty $e_g$ band just above the Fermi level,
 a highly conductive hopping path is provided.
 Besides, low thermal conductivities (compared with SrTiO$_3$ at optimal doping
 levels) below 1 W $\times$ m$^{-1}$ $\times$ K $^{-1}$ have been reported\cite{thermal1}
 in other related double-perovskite oxides,
 and these low values have been related to the double-perovskite structure 
 itself.
 This encourages to think that if they can be doped at an optimal level, 
 these systems could be potential candidates to become excellent TE materials.
 To test the validity of the previous hypothesis on the electronic structure, 
 we have carried out first-principles simulations combined with 
 semiclassical Boltzmann computations of the transport quantities.

 The rest of the paper is organized as follows.
 The requirements for the ideal double perovskite 
 are summarized in Sec.~\ref{sec:requirements}. 
 The methods utilized for the simulations are described in
 Sec.~\ref{sec:comp_details}.
 In Sec.~\ref{sec:results} we present the electronic
 structures, Seebeck coefficients, electrical conductivities and
 power factors of different double perovskites,
 both in the relaxed cubic structures and under biaxial strain,
 whose application has been already used  
 to modify the band structure and enhance the power factor in other
 materials.~\cite{la2nio4_1,la2nio4_2,bi2se3_strain_theo,bi2se3_strain_exp}
 We also present other methods to tune the band structure,
 based on cation intermixing or volume variations, and analyze a more realistic
 case with a fully relaxed structure. Finally, 
 we compare the results with the best $n$-type oxide 
 TE material up to date, SrTiO$_{3}$.
 
 \section{Selecting the candidates}
 \label{sec:requirements}

 The properties that the candidate double-perovskite materials
 should meet in order to achieve the ideal band structure
 with the $d^0/d^6$ filling scheme
 depicted in Fig.~\ref{fig:perovskite}(b) 
 can be enumerated as follows.
 (i) The cation on the B' site must be in a low-spin state to yield a filled 
 $t_{2g}$ shell. For this reason, transition metal elements from the 
 $4d$ and $5d$ rows are selected.
 (ii) The highly-degenerate electronic configuration at the 
 low electron-doping region 
 which provides the desired scheme for a high Seebeck coefficient
 requires an empty $d$-shell in the B-cation site.
 For the studied compounds, we choose only elements in their 
 most usual valence state, since these materials
 are ultimately expected to be grown in the laboratory. 
 (iii) We study only ordered double perovskites.
 This ordered structure is more likely to occur if
 there is a substantial difference in 
 size and/or valence between the B and B' cations. 
 To ensure that the cubic perovskite structure is
 preserved (or is at least a good approximation), 
 we also combine elements whose ionic radii
 fulfill the tolerance factor relation

 \begin{equation}
    t=\frac{r_{\rm A}+r_{\rm O}}
           {\sqrt{2}(\langle r_{\rm B} \rangle + r_{\rm O})}\lesssim 1,
    \label{eq:tolfac}
 \end{equation}

 \noindent where $\langle r_{\rm B} \rangle$ is the average ionic 
 radii of the B and B' cations, $r_{\rm O}$ is the oxygen's
 ionic radius and $r_{\rm A}$ is the A cation ionic radius. 
 If the tolerance factor is on the order of 1, the structure 
 tends to be more stable in the cubic phase.
 In principle, non-cubic perovskites would be useful as well, but a much more complex computational treatment is required due to distortions.

 Under these premises, we consider for the B site 
 Ti$^{4+}$, Nb$^{5+}$, and Ta$^{5+}$ cations 
 from the $3d$, $4d$, and $5d$ transition metal series, respectively.
 These are combined with Ir$^{3+}$, Rh$^{3+}$, Pt$^{4+}$, and Pd$^{4+}$
 cations for the B' site, 
 from the $4d$ and $5d$ transition metal series. 
 Typical A cations chosen are Sr$^{2+}$ and La$^{3+}$
 or an ordered mixture of both.

 \begin{figure*}[ht]
    \begin{center}
       \includegraphics[height=13.5cm]{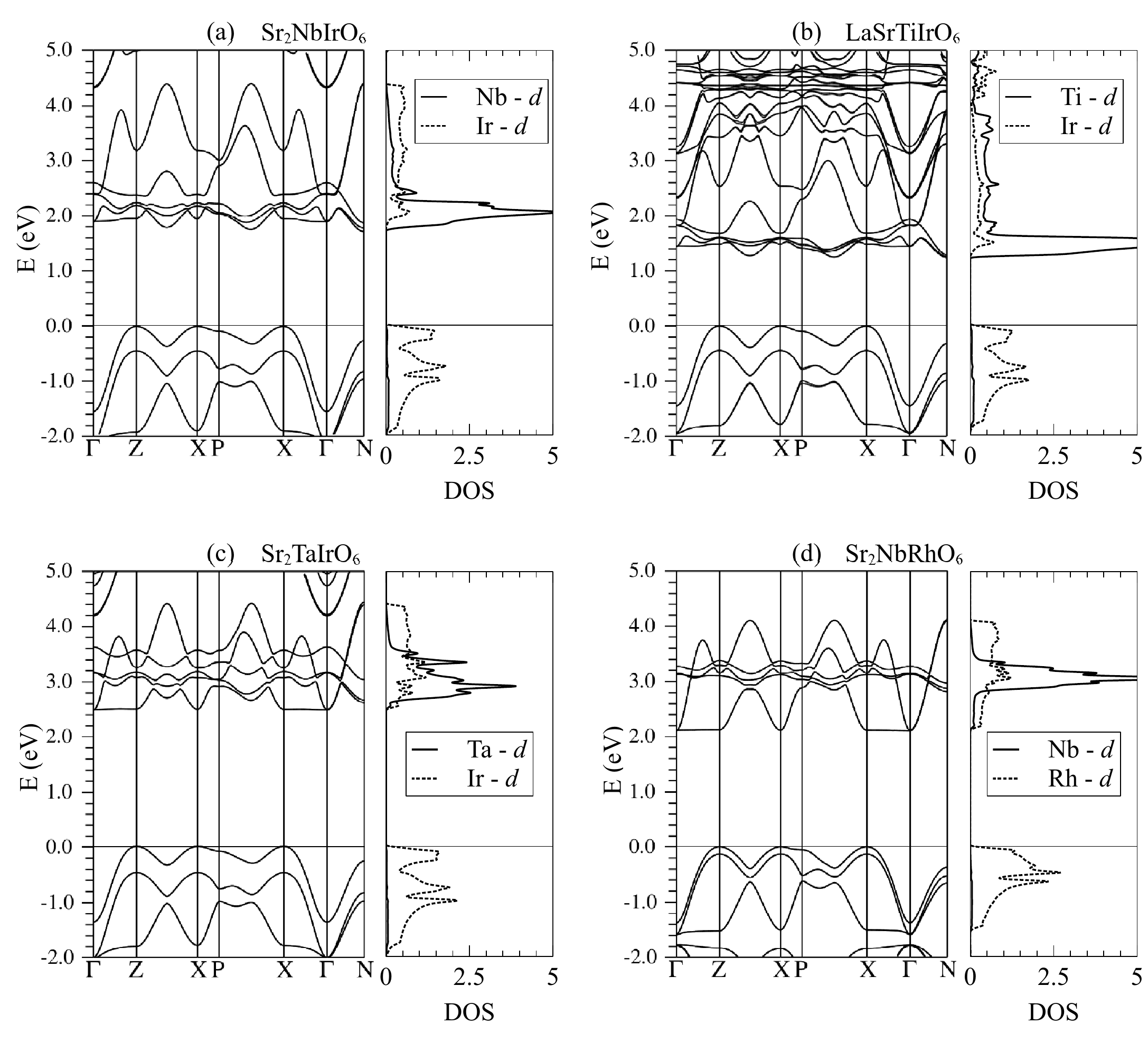}
       \caption{Band structure (left) and density of states (DOS, right) of 
                (a) Sr$_2$NbIrO$_6$, (b) LaSrTiIrO$_6$, (c) Sr$_2$TaIrO$_6$
                and (d) Sr$_2$NbRhO$_6$. 
                In the DOS plots, solid lines represent the projection on the 
                $d$-shell of the cations at the B-site (supposed to be in a 
                $d^0$ configuration),
                and dotted lines on the cations at the B'-site 
                (supposed to be in a $d^6$ configuration).
                The zero of the energy is set to the top of the valence bands 
                for all compounds and the scale is the same for all cases.
                Only the DOS projected on the $d$ electrons are represented  
                as they are 
                the main contributors to the bands that can be
                accessed by electron doping at the energy range considered here.
                Units: energies in eV and DOS in states/( eV $\times$ spin).}
       \label{fig:bandas}
    \end{center}
 \end{figure*}

\section{Computational details}
\label{sec:comp_details}

 Electronic band structures are calculated within the density functional theory
 (DFT)\cite{HK, KS} 
 framework using the all-electron, full potential code
 {\sc wien2k}.~\cite{WIEN2k}
 This package is based on the augmented plane waves plus local orbitals 
 (APW$+$lo) method.~\cite{sjostedt2000}

 Typical values for the muffin-tin spheres radii used are (in a.u.): 
 2.50 for Sr and La; 
 1.90 for Ta, Rh, Ir, Pd, Pt, Ti, and Nb;
 and 1.60 for O. 
 In some cases, especially under the most extreme strain conditions
 that are discussed in Sec.~\ref{sec:biaxial-strain}, 
 these radii were further reduced.

 Inside the muffin-tin spheres, the mixed APW+lo
 (where lo stands for local orbital) and LAPW basis set are
 expanded in spherical harmonics up to $\ell_{\rm max}^{\rm wf}=10$.
 Nonspherical contributions to the electron density and potential up to
 $\ell_{\rm max}^{\rm pot}=$ 6 are used.

 In LAPW-based methods, the plane-wave cutoff for the electronic wave functions
 in the interstitials is controlled by the parameter
 $R^{\rm MT}_{\min}\times$ K$_{\max}$,
 where $R^{\rm MT}_{\min}$ is the smallest muffin-tin sphere radius and
 K$_{\rm max}$ is the largest plane wave momentum vector defined by this
 product.
 A parameter $R_{\rm MT}\times$K$_{\max}$=7.0 ensures calculations are converged for the magnitudes presented, in particular optimal volumes and band gaps.

 The integrals in reciprocal space are well converged, using a sampling
 in $k$-space of $8 \times 8 \times 8$, generated using the modified tetrahedron method.\cite{tetra}
 For the computation of the density of states (DOS),
 a non-self-consistent calculation within a grid of
 $15 \times 15 \times 15$ is carried out.

 The relaxed lattice parameters are obtained within the 
 generalized gradient approximation (GGA),
 using the exchange and correlation functional proposed by 
 Wu and Cohen.~\cite{WCexcrrfun}
 This choice is motivated by the fact that it is known that 
 the Wu and Cohen functional improves the accuracy of the
 structural properties of bulk ABO$_{3}$ ferroelectrics over 
 the most usual GGA's.~\cite{Nishimatsu-10} 
 For each combination of cations, we performed an initial volume optimization 
 within a cubic perovskite structure, and the results
 on the TE properties are discussed in 
 Sec.~\ref{sec:relaxed}.
 Then, the effect of biaxial strain imposed by a hypothetical 
 substrate is simulated, constraining the length of the
 lattice vectors along the three cartesian directions
 in such a way that $a=b\neq c$.
 Assuming that the in-plane lattice constant, 
 $a$, is determined by the substrate, then
 the out-of-plane lattice constant $c$ is relaxed.
 The results are summarized in Sec.~\ref{sec:biaxial-strain}.
 The possibility of tuning the band structure
 under hydrostatic pressure is also explored (Sec.~\ref{sec:hydrostaticpressure}).
 In this case, volume reductions are performed decreasing 
 the cubic $a$ lattice parameter retaining the cubic symmetry. 
Finally, a full relaxation of the atomic structure is studied in Sec.~\ref{sec:full_relax},
for which all forces were relaxed below 2.0 mRy/a.u.

 Once the atomic structure is determined in every case, 
 we have used the Tran-Blaha modified Becke-Johnson (TB-mBJ) semilocal
 exchange potential for obtaining the band structures,~\cite{TB-mBJ}
 since this functional has been shown to 
 reproduce successfully the band gap for a wide range of 
 semiconductors and oxides,~\cite{TB-mBJ1, TB-mBJ2} SrTiO$_3$ among
 them.~\cite{elias_alex_sto}
 Due to the presence of heavy atoms with $4d$ and $5d$ electrons,
 spin-orbit interaction effects could be of significance.
 In order to assess the importance of spin-orbit coupling and
 eventual modifications of the band structure of these double perovskites,
 we have performed calculations including this effect in a second
 variational manner.~\cite{singhLAPW}

 An on-site Coulomb repulsion $U$ for the compounds containing La 
 is used to move La-$4f$ empty levels~\cite{anisimov1997first, gillen2012nature}
 away from the Fermi energy.
 This is motivated by the fact that these orbitals do not participate 
 in the calculation of the transport coefficients,~\cite{boltztrap, transcoeff}
 which involves only a very narrow energy range
 related with the $d$-bands,
 and its explicit consideration might introduce spurious contributions in the
 TE phenomena analyzed.
 This is carried out with the usual LDA+$U$\cite{sic1,fll} prescription
 (with an $U$ around 9 eV),
 where the uncorrelated part of the exchange-correlation 
 is obtained using the TB-mBJ scheme explained above.
 Realistic inclusion of 4$f$ orbitals would require to adequately 
 select $U$ to match the experimental position of these 
 levels in the studied double perovskites,
 but this is beyond the scope of the current work.

 We have estimated the electrical conductivity and the Seebeck coefficient
 through the semiclassical Boltzmann theory~\cite{Ziman, Ashcroft}
 within the constant relaxation time approximation, 
 as implemented in the {\sc BoltzTraP} code.~\cite{boltztrap}
 This implementation relies on the Fourier expansion of the band-energies,
 provided by a first-principles electronic structure code 
 ({\sc wien2k} in the present study).
 In this case, a denser $k$-mesh of $25\times25\times25$ is needed 
 to reach convergence for the Fermi-surface integrals involved in the 
 TE quantities presented. Since we are using the simple constant relaxation time 
 approximation, the scope of our paper is to provide on the one hand trends with strain and 
 also a relationship between electronic structure and transport properties. It is not our intent
 to give accurate values for the optimal doping or Seebeck coefficients, these
 should be taken with caution when comparing with future experiments.

\section{Results and discussion}
\label{sec:results}

\subsection{Relaxed case electronic structure}
\label{sec:relaxed}

 Figure~\ref{fig:bandas} shows the band structures and densities of states (DOS) 
 for some representative double-perovskite cases. 
 It is interesting to compare their basic electronic structure with the 
 rough picture described above as the desired electronic structure 
 in the cartoon of Fig.~\ref{fig:perovskite}(b).
 For Sr$_2$NbIrO$_6$, Fig.~\ref{fig:bandas}(a), 
 narrow Nb $t_{2g}$ bands are just at the bottom of the conduction band,
 almost degenerate with the wider Ir $e_g$ bands. 
 The same happens with other compounds
 like LaSrTiIrO$_6$ [Fig.~\ref{fig:bandas}(b)] 
 or Sr$_2$TaIrO$_6$ [Fig.~\ref{fig:bandas}(c)],
 all of them characterized by the presence of a $5d$ cation, Ir,
 at the B'-site.
 However, there are some other cases like Sr$_2$NbRhO$_6$ 
 [Fig.~\ref{fig:bandas}(d)] which present a
 slightly shifted band structure, with the unoccupied 
 $t_{2g}$ bands of the $d^0$ metal lying at
 the middle of the B'-$e_g$ bands. 
 All in all, the main result that can be drawn is that these double perovskites
 combine a localized B-$t_{2g}$ band and a more delocalized B'-$e_g$ band 
 that lie roughly in the same energy window. 
 The actual details of band orderings depend on the particular
 atomic configuration of the system but,
 to a first approximation, we find that it is possible 
 to obtain an electronic structure similar to the one sketched in Fig.~\ref{fig:perovskite}(b).

 \begin{figure}[!ht]
    \begin{center}
       \includegraphics[height=0.9\columnwidth]{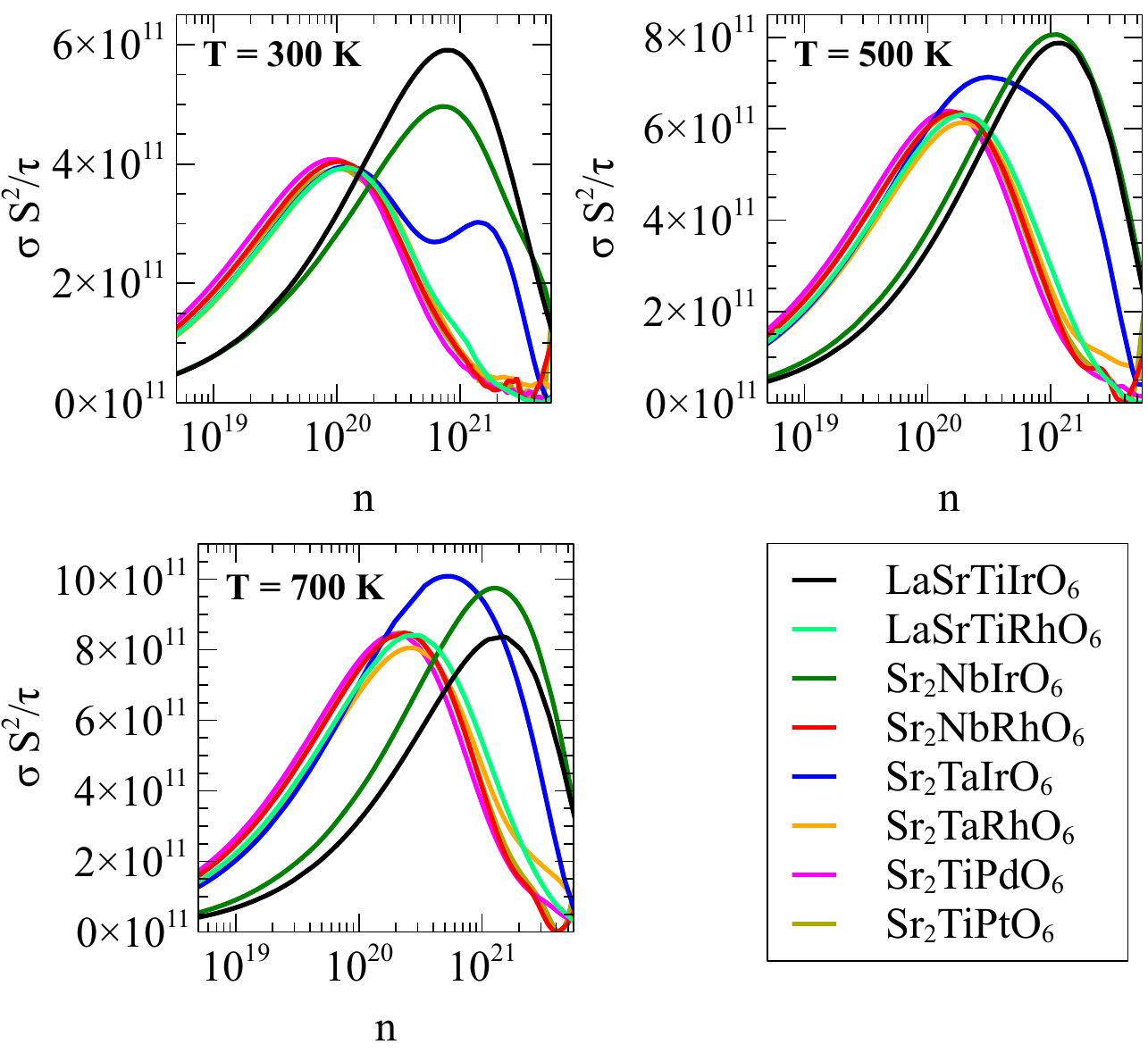}
       \caption{(Color online) Power factor divided by the scattering time
                for all studied compounds at three different temperatures 
                as a function of carrier concentration.
                Units: power factor over relaxation time in W/(mK$^{-2}$s), 
                carrier densities in cm$^{-3}$. }
       \label{fig:pf}
    \end{center}
 \end{figure} 

 \begin{figure*}[ht]
    \begin{center}
       \includegraphics[height=16cm]{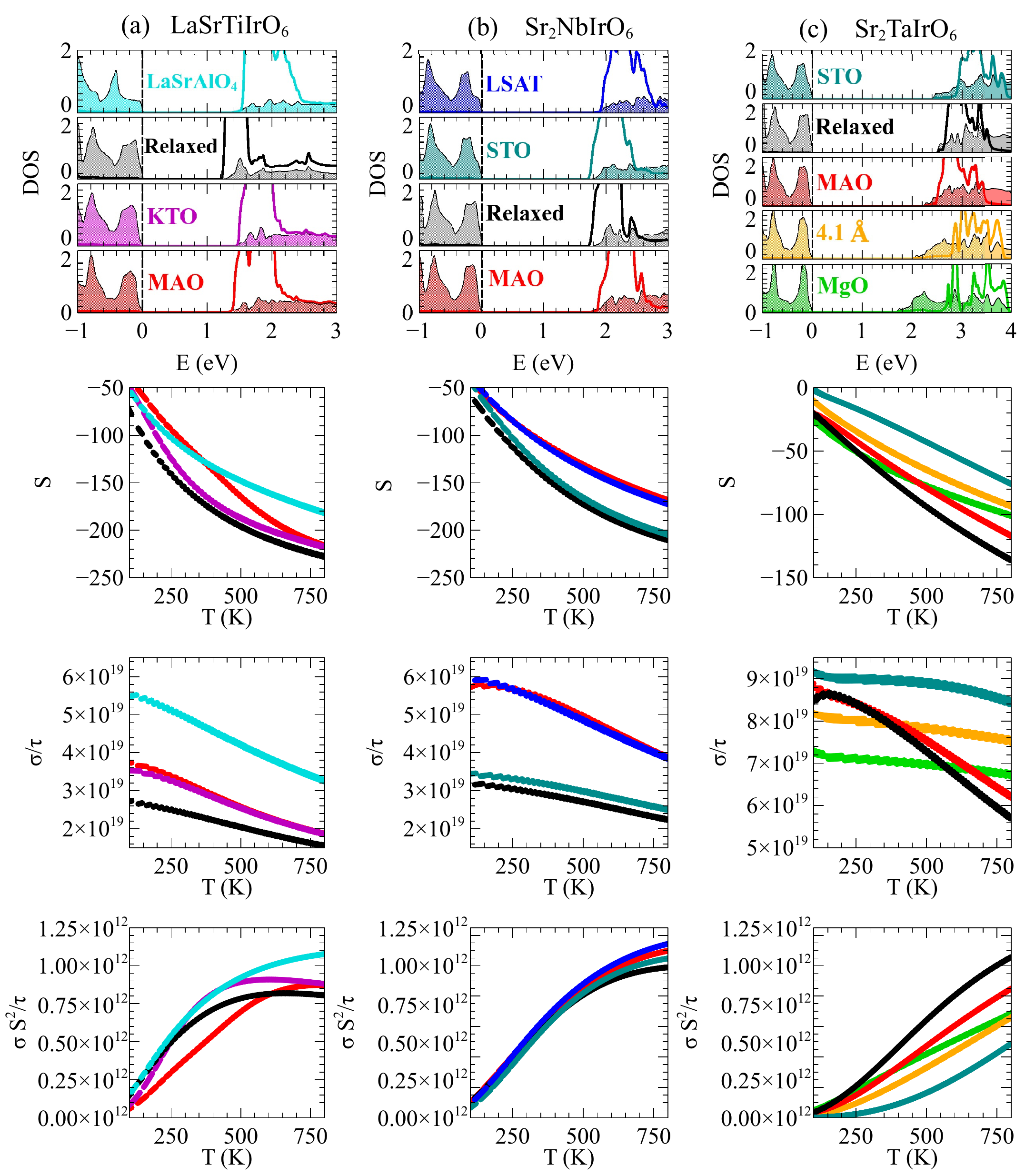}
       \caption{(Color online). 
                Density of states [DOS, in states/(eV $\times$ spin), top row], 
                and temperature dependence of the 
                Seebeck coefficient ($S$, in $\mu$V/K, second row),
                electrical conductivity over relaxation time
                [$\sigma/\tau$, in 1/($\Omega$ $\times$ m $\times$ s), 
                third row],
                and power factor over relaxation time
                [$\sigma S^{2} / \tau$, in W/(m $\times$ K $^2\times$ s), 
                bottom row] of 
                (a) LaSrTiIrO$_6$, (b) Sr$_2$NbIrO$_6$ and (c) Sr$_2$TaIrO$_6$.
                Different strains have been simulated by setting the 
                lattice parameter $a$ to the experimental value of 
                LaSrAlO$_{4}$ (light blue),
                LSAT (dark blue),
                STO  (dark green),
                KTO  (magenta),
                MAO  (red),
                MgO  (light green),
                and with a value of $a$ = 4.10 \AA (yellow).
                DOS for the relaxed lattice constant
                within the cubic symmetry of each compound are 
                shown in grey.
                Shaded DOS corresponds to the projection on the $d$-shell of the
                B'-cation (Ir in the presented examples, 
                assumed to be in a $d^6$ configuration).
                Projections on the $d$-shell of the B-cations
                (narrower $t_{2g}$ bands) are represented by a solid line.
                DOS panels over (below) the relaxed plot means 
                compressive (expansive) epitaxial strain.
                The zero of the energy is set to the top of the valence bands.}
       \label{fig:test}
    \end{center}
 \end{figure*}

 In order to analyze the TE properties of these double perovskites,
 we show in Fig.~\ref{fig:pf} the power factor divided by the relaxation time
 as a function of the carrier concentration
 for three different temperatures.
 The calculations are performed within the semiclassical Boltzmann theory with 
 two approximations:
 (i) the constant relaxation time, and 
 (ii) the ``rigid band approach'' that assumes that the band structure
 does not change with temperature or doping and, therefore, it is fixed
 independently of the chemical potential.
Despite the effect that the energy, temperature or doping dependence of the relaxation time may have on the final
figure of merit of the material,~\cite{relaxation_time} we want to stress in this point the importance of having 
an appropriate electronic structure which can be easily tuned by different means, as it happens in these compounds.
  The values of $\sigma$ and thermopower presented are the average value of the diagonal components of the TE tensors. Due to the cubic symmetry of the system, this is a very representative quantity.
 The main results are:
 (i) We observe that for all compounds including Ir, the optimal carrier concentration
 for a maximal power factor and optimal TE performance is 
 approximately $n=10^{21}$ cm$^{-3}$, while for those double perovskites 
 which do not include Ir, the maximum
 is at lower carrier concentrations around $n=10^{20}$ cm$^{-3}$.
 (ii) These optimal doping levels are almost temperature independent;
 only small deviations of the order of 10\% are observed as temperature 
 is increased.
 (iii) The maxima of the power factor at optimal doping are substantially 
 enhanced with temperature, increasing by about 70\% from 300 K to 700 K.
 The origin for this behaviour can be traced back to 
 the band structures shown in Fig.~\ref{fig:bandas},
 where it can be seen how the Ir-based
 compounds present a band degeneracy at the bottom of the conduction band formed
 by the Ir $e_g$
 and Ti/Nb/Ta $t_{2g}$ bands. 
 Such degeneracy is intimately related
 to an enlarged power factor and a higher optimal carrier concentration. 
 The relationship between large band valley degeneracy and enhancement 
 of the TE properties is well known,\cite{valley_degeneracy} and our results 
 confirm that this is possible to obtain with these double perovskites 
 to some extent.

\subsection{Tuning the band structure by biaxial strain in the ab-plane}
\label{sec:biaxial-strain}

 It has been demonstrated 
 theoretically~\cite{Hicks-93.1,Hicks-93.2,ohta2007giant} 
 and also tested experimentally~\cite{boukai2008silicon,hochbaum2008enhanced} 
 how quantum confinement and low-dimensionality could enhance the TE figure 
 of merit of a system. 
 Also, strain has been found to play a role in decoupling the 
 different transport 
 coefficients involved in TE efficiency, which are usually
 inter-related.~\cite{la2nio4_1,la2nio4_2,bi2se3_strain_theo,bi2se3_strain_exp} 
 Thin films of similar double perovskites
 have been grown routinely over different
 substrates.~\cite{guo2006growth, manako1999epitaxial} 
 Thus, it is plausible to explore the effects of biaxial strain 
 (which is achieved by means of epitaxial growth of thin
 films over substrates with a different lattice parameter) on these compounds. 
 We can theoretically analyze the effects of both compressive
 and tensile strain in the band structure of these materials by simply 
 imposing the in-plane $a$
 lattice parameter and relaxing the $c$ out-of-plane lattice constant keeping
 the constraint of a tetragonal symmetry.
 By studying the general changes in the electronic structure and TE properties 
 we can try to generalize
 and construct a recipe to improve the TE response of these double perovskites.

 For a better comparison with experiments, we have selected the $a$
 lattice constant
 corresponding to the experimental value of different commonly used substrates,
 that cover a wide range of $a$ lattice constants: 
 LaSrAlO$_4$ ($a_{\rm{LaSrAlO_4}}$= 3.755 \AA), 
 (LaAlO$_{3}$)$_{0.3}$(Sr$_{2}$TaAlO$_{6}$)$_{0.7}$ 
 (LSAT, $a_{\rm LSAT}$ = 3.868 \AA), 
 SrTiO$_3$ (STO, $a_{\rm STO}$ = 3.905 \AA), 
 KTaO$_3$ (KTO, $a_{\rm KTO}$ = 3.989 \AA), 
 MgAl$_2$O$_4$ (MAO, $a_{\rm MAO}$ = 4.041 \AA),
 MgO ($a_{\rm MgO}$ = 4.212 \AA),
 and another set of calculations with $a=4.1$ \AA\ 
 to fill the gap between MAO and MgO.

 The epitaxial strain changes both the position of the center 
 and the widths of the bottom conduction bands in systems
 with octahedral symmetries, such as the double perovskites 
 considered here.
 For $t_{2g}$ bands coming from the B cation,
 on the one hand, a compressive epitaxial strain produces an elongation of the 
 oxygen octahedra along the $c$ direction that stabilizes the $d$ orbitals
 with $z$ components, so the center of the respective bands 
 (the $d_{xz}$ and $d_{yz}$ doublet) are positioned
 lower in energy than the $d_{xy}$ singlet.
 Moreover, compressive strain reduces the distance
 between the atoms in the $ab$ plane and produces larger overlap between the
 orbitals directed in the plane perpendicular to the strain axis
 ($d_{xy}$ and $d_{x^{2}-y^{2}}$), so the width of the corresponding
 bands increases.
 On the other hand, a tensile epitaxial strain produces a compression of 
 the octahedra along the $c$-direction, that will place the $d$ orbitals
 with $z$ components higher in energy.~\cite{Valionis-15} 
 In this case, the bands with major character from orbitals 
 with $z$ component are wider.

 A similar effect occurs for the B'-$e_g$ bands: tensile (compressive) strain 
 lowers the $d_{x^2-y^2}$ ($d_{z^2}$)
 band and band widths get enlarged for the in-plane (out-of-plane) orbital 
 when compressive (tensile) strain is applied.
 All these effects combined lead to different band displacements that move in 
 different directions when epitaxial strains are applied. 
 The main results are summarized for selected compounds in Fig.~\ref{fig:test}.

 In those cases in which the involved B'-$e_g$ manifold 
 is lower in energy than the B-$t_{2g}$ manifold, 
 such as LaSrTiIrO$_6$ over LaSrAlO$_4$ substrate [Fig.~\ref{fig:test}(a)],
 or Sr$_2$NbIrO$_6$ over LSAT or MAO substrates [Fig.~\ref{fig:test}(b)]
 no drastic effect
 is observed in the Seebeck coefficient by applying strain
 (sometimes it gets even slightly reduced because of the wider B-$t_{2g}$ bands).
 However electrical conductivity in those cases increases to even 2 times
 the value for the relaxed case within the cubic symmetry
 (from $2.2\times10^{19}$ to $4.0\times10^{19}$ 1/($\Omega$ $\times$ m $\times$ s) at 800 K for Sr$_2$NbIrO$_6$ [Fig.~\ref{fig:test}(b)]).
 The result is an enhanced power factor 
 that takes place together with a large degeneracy at the 
 bottom of the conduction band.
 We present these TE properties
 obtained at $n=10^{21}$ cm$^{-3}$
 for the three materials appearing in Fig.~\ref{fig:test}.
 Depending on the compound, the optimal carrier concentration could vary, but trends are consistent in the 
 10$^{18}$ - 10$^{21}$ cm$^{-3}$ range. It is not the objective of this work to provide accurate values of the optimal doping. 
 Band structure tuning by shifting bands 
 and modifying their band widths 
 by applying strain is most effective when the bottom of the 
 B-$t_{2g}$ and B'-$e_g$
 bands get close in energy, as can be clearly seen 
 for LaSrTiIrO$_6$   [Fig.~\ref{fig:test}(a)]
 and Sr$_2$NbIrO$_6$ [Fig.~\ref{fig:test}(b)].
 One can also notice in these two panels that 
 if the B-$t_{2g}$ band gets occupied first
 (at this electron-doping level we are considering) 
 the electrical conductivity gets drastically reduced and thus the power factor is lower 
 [in Fig.~\ref{fig:test}(a) for all substrates except for LaSrAlO$_4$ and in 
 Fig.~\ref{fig:test}(b) for STO and the relaxed case within the cubic symmetry].

 In Fig.~\ref{fig:test}(c) we see the results for Sr$_2$TaIrO$_6$,
 which confirms the trend that the presence of a lower lying B'-$e_g$ 
 band is necessary
 for an enlarged power factor to occur. 
 The other important fact seen here is that the wider the $t_{2g}$ bands, 
 the lower the Seebeck coefficient is,
 which is not a surprising effect.
 In some cases, even
 biaxial strain does not enhance the TE properties since the 
 relaxed case within the cubic symmetry 
 already presents an optimal configuration.

 For the other compounds (not shown) strain does not play a major role. 
 In the limit in which the B-$t_{2g}$ bands lie in the middle of the 
 broader B'-$e_g$ bands, the effect is very small because it does not 
 largely affect the doping region
 where large band valley degeneracy occurs. 
 The value of the power factor obtained for the corresponding
 double perovskites is lower
 than in the other compounds because the Seebeck coefficient 
 is not so large.
 Even when biaxial strain is applied, we do not observe a 
 significant improvement of the TE properties.

 We conclude that an initial configuration with the B-$t_{2g}$ and B'-$e_g$ 
 bands being both at the bottom of
 the conduction band is required for having good TE properties in 
 these systems, and that in those cases it is possible
 to tune them by means of biaxial strain.

\subsection{Other methods to tune the band structure}
\label{sec:othermethods}

 Besides the application of biaxial strains imposed by epitaxial growth,
 we have examined three other mechanisms to modify
 the electronic band structure in order to enhance the TE performance:
 the effect of using more compact $d$-states considering atoms of the 3$d$-row,
 the use of hydrostatic pressure to reduce the unit-cell volume and change the
 distance between atoms, and chemical substitution to induce a 
 ``chemical pressure" effect.
 The main results are explained in the following three subsections.

 \subsubsection{Spatial range of the $d$-orbitals}
 \label{sec:extension-d}

 The effect of the extension of the $d$ orbitals of the B- and B'-cations and 
 the corresponding change in hybridizations
 has been studied by replacing the previously considered atoms  
 in the $4d/5d$-row by transition metal atoms of the $3d$ row.
 For this purpose we have chosen La$_2$TiFeO$_6$. 
 The existence of ordered cationic structures in this compound can be 
 presumed by the large valence difference. 
 The calculation converges nicely to a Ti$^{4+}$($d^0$)/Fe$^{2+}$($d^6$),
 where Fe$^{2+}$ is in a low-spin state with a full $t_{2g}$ shell, 
 just like every other compound we have examined previously in this study. 
 We immediately observe that even though the Seebeck coefficient (not shown) remains as
 large as for the other double perovskites presented, 
 the electrical conductivity is at least one order of
 magnitude lower than the previous cases. 
 The driving force responsible for this reduction can be ascribed to the more localized character of Fe $3d$-orbitals compared with $4d$ or $5d$ orbitals.
 The compactness of the $d$-orbitals makes the hopping between Fe and O lower 
 than the corresponding hopping due to a $4d$/$5d$ metal. 
 The band structure (not shown) presents the Fe $e_g$
 bands above the Ti $t_{2g}$ bands, so the conduction mediated by the $e_g$ bands 
 (which worked in other compounds) is not active in this case, and thus 
 the electrical conductivity is reduced.
 For these reasons, we have not considered further combinations with $3d^0/3d^6$ transition
 metal cations and focused only on the combination of $4d/5d$ elements at
 the B- and B'-sites.

 \subsubsection{Hydrostatic pressure}
 \label{sec:hydrostaticpressure}

 \begin{figure}[!ht]
    \begin{center}
       \includegraphics[height=\columnwidth]{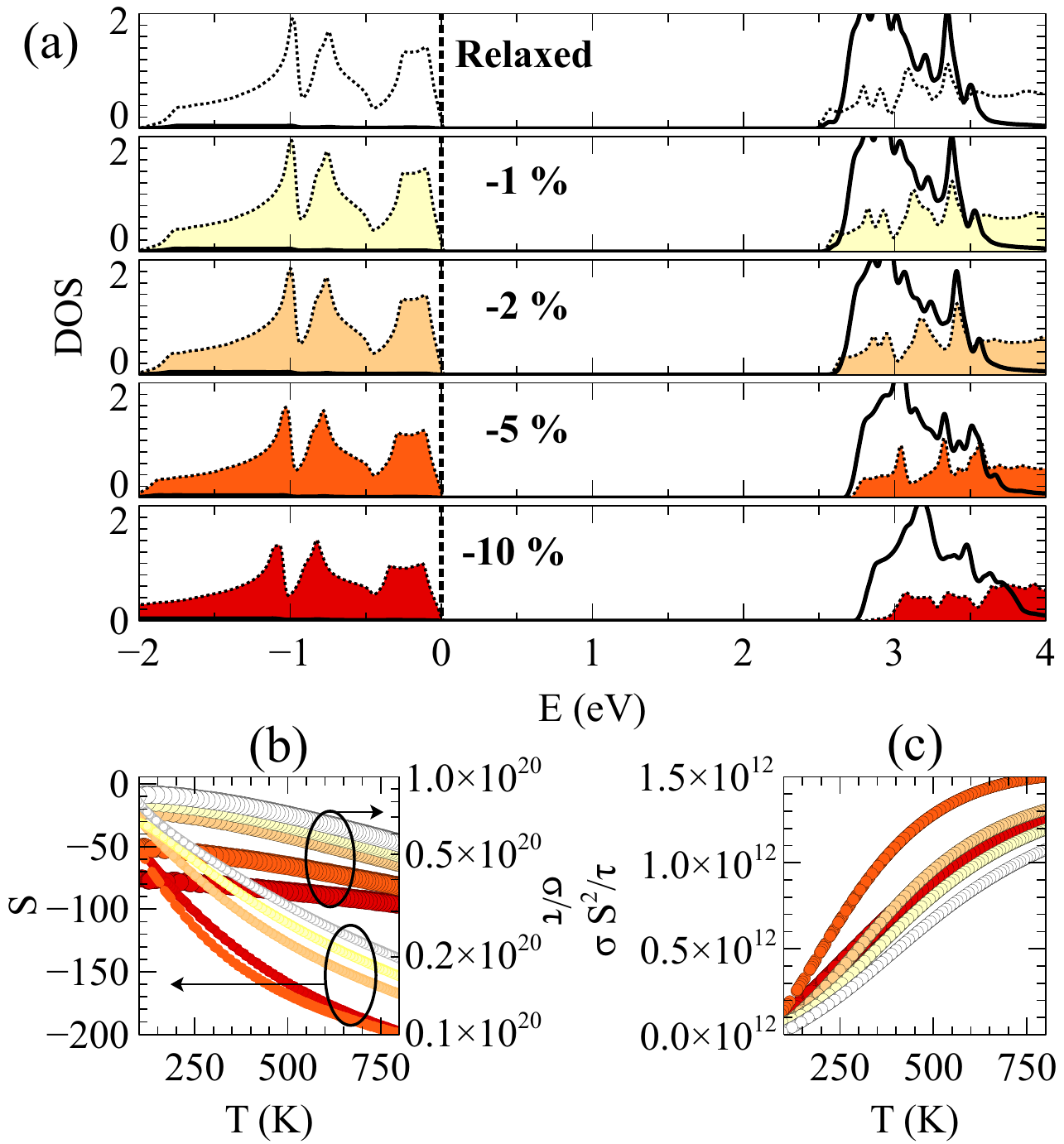}
       \caption{(Color online) Density of states, thermopower,
                electrical conductivity and power factor of Sr$_2$TaIrO$_6$ 
                at different unit-cell volumes.
                The color sequence follows from white for the relaxed 
                case in the absence of pressure, to red for the case 
                under the highest pressure (larger volume reduction).
                Corresponding values of volume reductions are
                indicated on the top panel.
                Meaning of lines, symbols and units as in Fig.~\ref{fig:test}. The electrical conductivity divided by the relaxation time 
                is represented in logarithmic scale.}
       \label{fig:volume}
    \end{center}
 \end{figure}

 We have also studied the possibility of enhancing the TE properties by means of
 the application of an external hydrostatic pressure.
 The most important effect induced by pressure is the reduction 
 of the unit cell volume, that is assumed to remain in a cubic symmetry. 
 The structural changes and the concomitant smaller metal-oxygen distances
 modify the electronic band structure of the double perovskites accordingly. 
 In particular, the crystal field
 splitting between $t_{2g}$ and $e_{g}$ levels will be enhanced as 
 pressure is increased.
 This translates into a lower positioning
 of the B-$t_{2g}$ bands with respect to the B'-$e_g$ bands that are 
 shifted upwards in the conduction band.

 As a test case, we have studied the effect of hydrostatic pressure on
 Sr$_2$TaIrO$_6$, with volume reductions up to 10 \% which would correspond to 
 external pressures of about 27 GPa.
 In this double perovskite, at the relaxed cubic structure in the 
 absence of external
 pressure, the bottom of the narrow Ta $t_{2g}$ bands are located 
 at the same energy as 
 the bottom of the more widely spread Ir $e_g$ manifold [see DOS at the top of 
 Fig.~\ref{fig:volume}(a)].
 Thus, it is a good example where the evolution of these bands with 
 volume variation can be studied.
 When pressure is applied and the unit cell volume is reduced, 
 the crystal field gap between the $t_{2g}$ and the $e_{g}$ bands for each atom 
 becomes larger, as it happens also with the width of each of these bands
 [Fig.~\ref{fig:volume}(a)].
 However, both variations are not dramatic, 
 significant changes are only expected if the initial positioning 
 of the bottom of the $t_{2g}$ and $e_g$ bands is really close. 

 Regarding the transport properties [Fig.~\ref{fig:volume}(b)]:
 on the one hand, the Seebeck coefficient is enlarged (in absolute value) 
 when the degeneracy at the bottom of the
 conduction band is maximal.
 On the other hand, the electrical conductivity is reduced when the 
 volume decreases,
 i.e. when the system tends to have the more localized
 Ta $t_{2g}$ bands occupied first, instead of the more spread Ir $e_g$ ones. 
 Combining all these effects [Fig.~\ref{fig:volume}(c)], 
 the higher power factor occurs for a volume reduction
 of 5\% in the unit cell, equivalent to a pressure of 11 GPa.

 Again, the final conclusion is that if the involved bands are already 
 close to the optimal band structure configuration,
 an enhancement of the TE response can be achieved by means of a
 reasonable volume reduction.

 \subsubsection{Chemical pressure}
 \label{sec:chemicalpressure}

 \begin{figure}[!ht]
    \begin{center}
       \includegraphics[height=6.1cm]{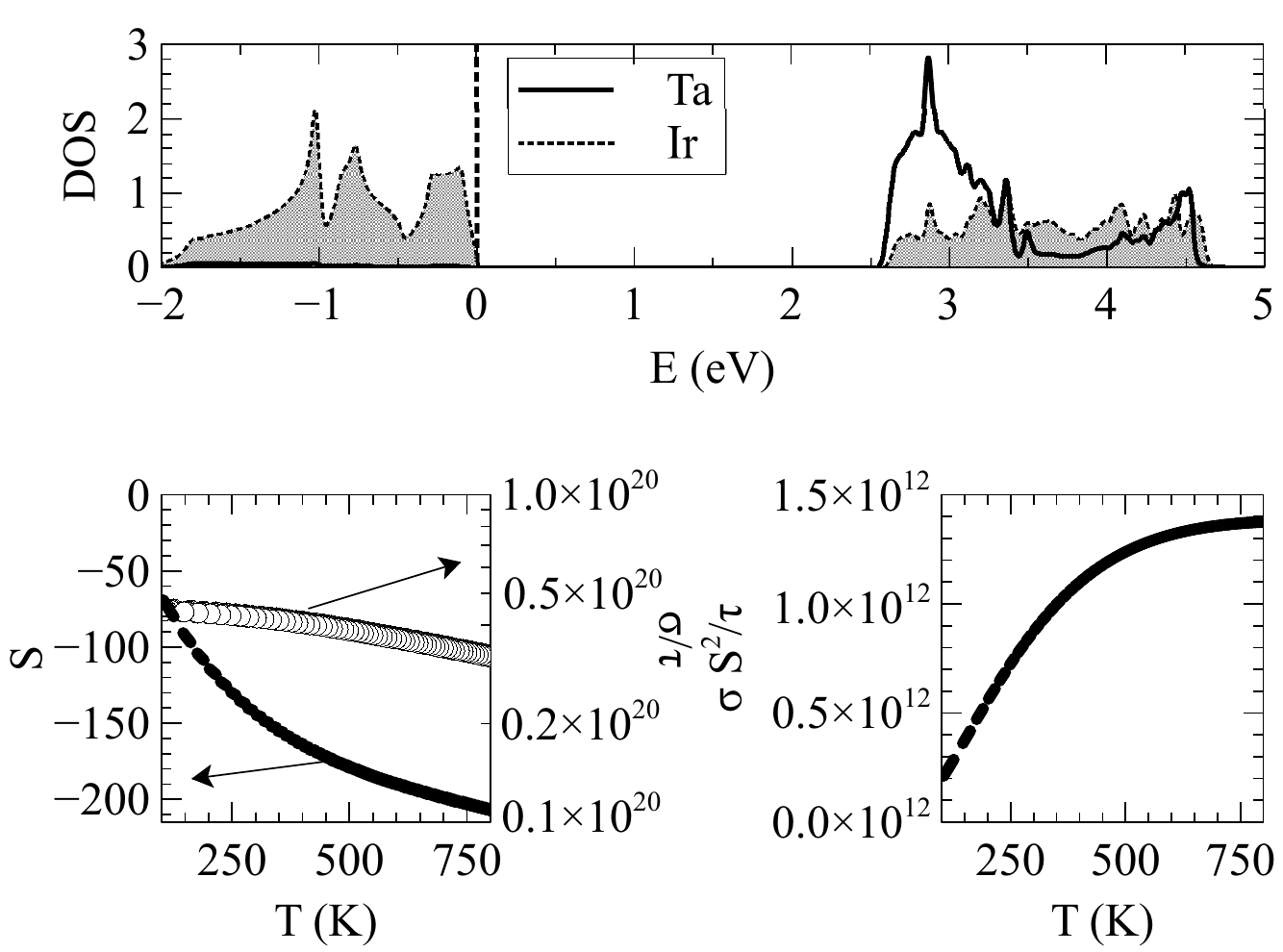}
       \caption{ Electronic structure and thermoelectric properties
                 of Ca$_2$TaIrO$_6$. When compared to Sr$_2$TaIrO$_6$ under pressure, the tendency towards enhanced thermoelectric properties at lower volumes is retained. 
                 Meaning of lines, symbols and units as in Fig.~\ref{fig:test}. The electrical conductivity divided by the relaxation time 
                is represented in logarithmic scale.}
       \label{fig:ca2tairo6}
    \end{center}
 \end{figure}

 Volume reductions can also be realized by changing the size of the
 A cation (so-called chemical pressure).
 We have explored this possibility replacing 
 Sr$^{2+}$ (ionic radius: 1.44 \AA\ ) by smaller
 Ca$^{2+}$ (ionic radius: 1.34 \AA\ ) in the double perovskite
 studied in the previous subsection.  
 This substitution immediately reduces the lattice parameter of the 
 hypothetical cubic phase, that changes from 
 3.975 \AA\ in Sr$_2$TaIrO$_6$ to
 3.939 \AA\ in Ca$_2$TaIrO$_6$ (volume reduction of 2.7 \%).

 If we look at the electronic structure and transport properties of 
 Ca$_2$TaIrO$_6$ (Fig.~\ref{fig:ca2tairo6})
 we can see that the values for the Seebeck,
 electrical conductivity, and power factor are essentially the same as
 the ones for Sr$_2$TaIrO$_6$ with a 5\% volume reduction.
 In other words, the chemical substitution (even without applying any external
 pressure) produces the same effects as 
 the unit cell shrinking that optimized the TE figure of merit.

 The main drawback of this approach is that the theoretical tolerance factor
 [Eq.~(\ref{eq:tolfac})] is $t=0.948$ for this compound, 
 so we cannot assume that cubic symmetry will be preserved.
 Therefore, we acknowledge that these particular calculations could be 
 overestimating the real values for the TE properties.
 More complete and computationally demanding simulations 
 would be required to check these results, 
 involving a full relaxation of the lattice and atomic positions.

\subsection{Full relaxation}
\label{sec:full_relax} 

 For completeness, we have performed a full relaxation of the double-perovskite structure for one particular case to study its possible effects,
 even though a comprehensive study is beyond the scope of this paper. We have studied the compound
Sr$_2$NbIrO$_6$ under different strains. It presents an electronic structure close to the ideal configuration and a tolerance factor of $t=0.98$. Tilted octahedra
result from this structural relaxation. This influences directly the electronic structure and as a consequence, the
TE properties of the system. Even though the unit cell preserves a global cubic symmetry, the local environments for the cations become tetragonally distorted.
 
\begin{figure}[!ht]
\begin{center}
\includegraphics[height=10cm]{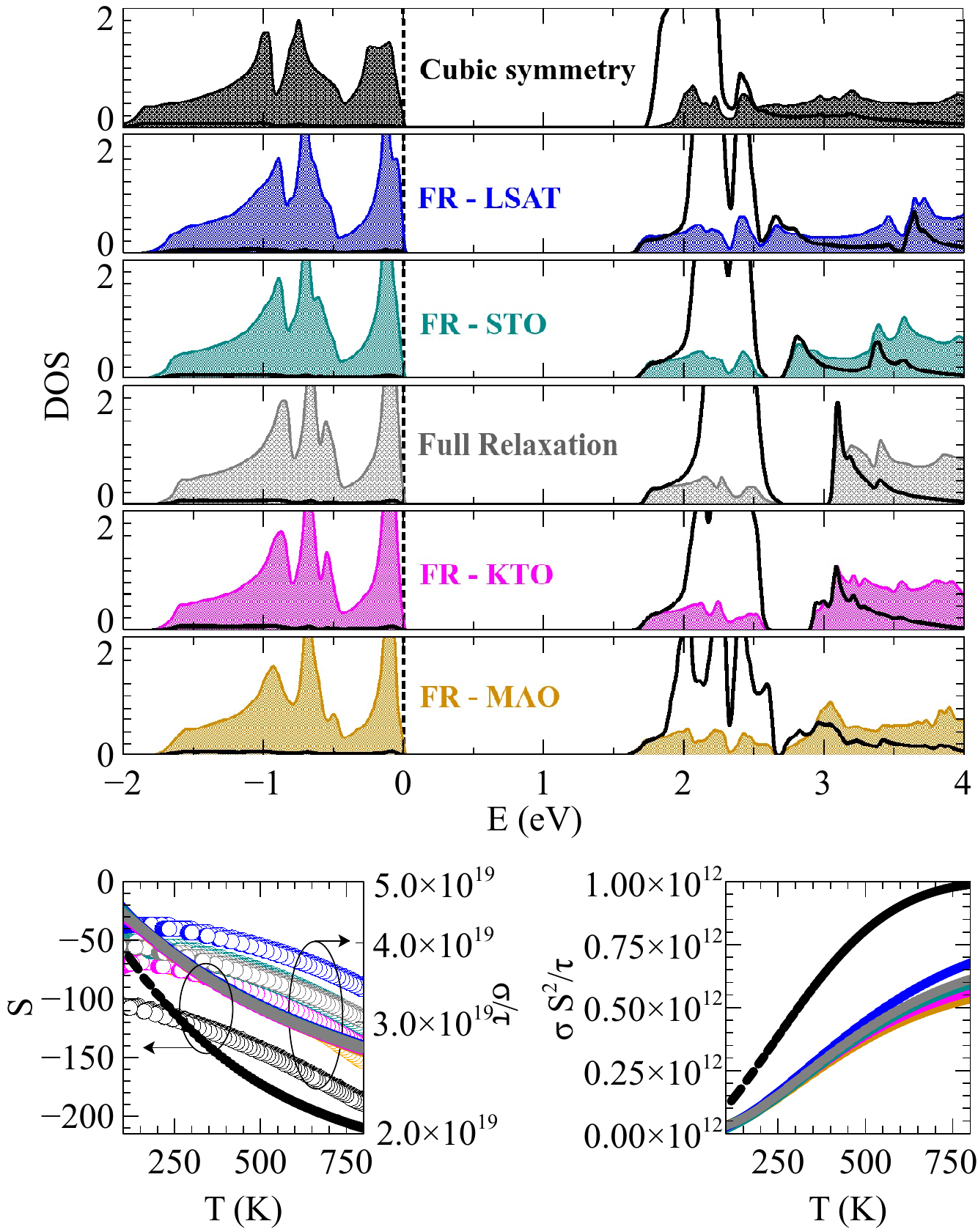}
\caption{Electronic structure and thermoelectric properties of Sr$_2$NbIrO$_6$ for the cubic structure (black), for the
fully relaxed structure (grey), and for the fully relaxed structure plus strain in the $ab$-plane (colors). In both cases, solid black lines represent the Nb $d$ bands, 
meanwhile the shaded density of states represents the Ir $d$ bands.}
\label{relaxation}
\end{center}
\end{figure}

We can see these differences in Fig. \ref{relaxation}, in which a comparison between the cubic symmetry case
and the fully-relaxed case under different strains is shown. Comparing the cubic-symmetry case with the fully relaxed case without strain, 
the most notorious effect is the splitting of the Ir $e_g$ manifold.
The degeneracy existing within these levels as a consequence of the local tetragonal symmetry is lifted by octahedral rotations. 
This directly affects the electrical conductivity of the system, since the hopping channel
is substantially modified. 

For all cases containing a full relaxation, a lower (and almost similar) Seebeck coefficient is present. This is due to the 
always low-lying Ir $e_g$ which, as we previously discussed, leads to a reduced TE power.
A higher electrical conductivity happens in the case of a fully relaxed structure with
the substrate LSAT (data in blue in Fig. \ref{relaxation}). The $e_g$ bands of Ir
(which in this case are not split) provide a wider hopping channel since they are lower in energy than 
the Nb $t_{2g}$ bands. Comparing this case with the one with cubic symmetry which also presents
degenerate $e_g$ bands (data in black in Fig. \ref{relaxation}) one can see the 
difference between the Seebeck coefficient and the electrical conductivity for both cases.
The difference in the electrical conductivity is explained as before,
taking into account which hopping path is activated first: the Ir $e_g$ bands in the case of a fully relaxed structure with LSAT
and the Nb $t_{2g}$ bands in the case of cubic symmetry. This results in a higher electrical conductivity in the first case. However the Seebeck coefficient is
lower in the case of the LSAT substrate because there is no peak in the density of states close to the bottom of the conduction bands, which happens in the case
of cubic symmetry. 

We are able to obtain a power factor for a fully relaxed structure on the order of 65-70\% the value of the power
factor in the simplified cubic case. With this result, we want to stress that if one wants to compute the electronic
structure of a real compound in a double-perovskite structure, it is necessary to include
a full relaxation of the structure, since the transport coefficients are quite sensitive to small changes in
the atomic positions. However, the main picture on how the ideal electronic structure would look like does not change, and the main effects discussed throughout the text remain the same.
Other calculations in similar systems involving full structural relaxations also retain a large TE efficiency.\cite{relaxations}

\subsection{Comparison with SrTiO$_3$}
\label{sec:srtio3}

 \begin{figure}[!ht]
    \begin{center}
       \includegraphics[height=3.2cm]{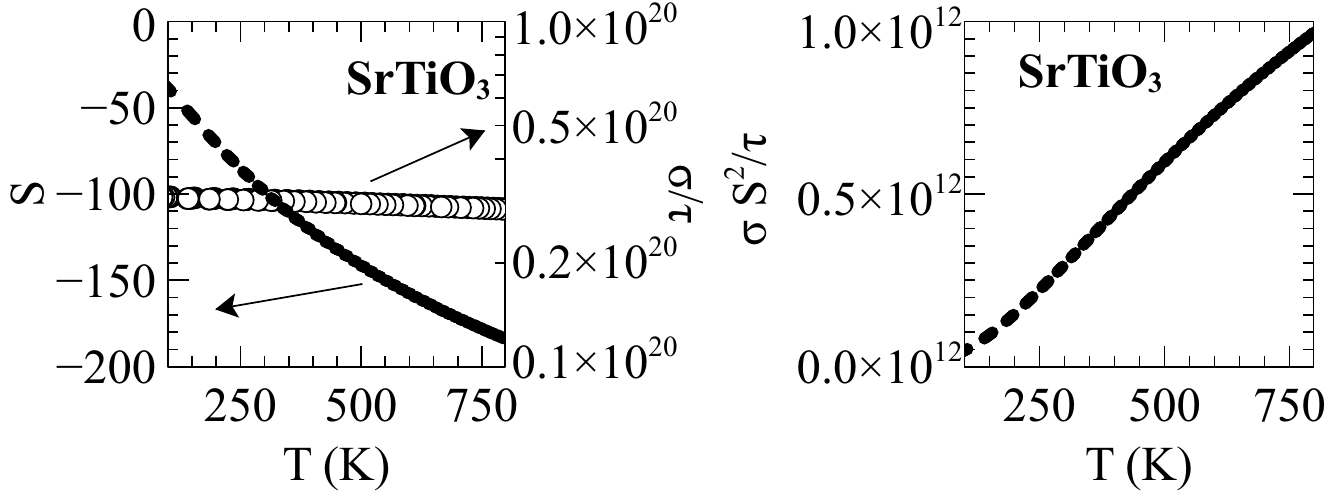}
       \caption{Thermoelectric properties for SrTiO$_3$.
                The electrical conductivity divided by the relaxation time 
                is represented by bigger circles and its corresponding axis
                is in logarithmic scale. Units as in previous figures.}
       \label{fig:STO}
    \end{center}
 \end{figure} 

 In order to gauge the importance of the enhancements of the TE performance
 obtained with the tuning of the band structure mechanisms explained in 
 Sec.~\ref{sec:results}, 
 we have carried out calculations for bulk SrTiO$_3$ (STO),
 one of the best $n$-type oxide TE materials found 
 up-to-date.~\cite{ohta2007giant,elias_alex_sto,heavydopping1,heavydopping2,heavydopping3}
 Figure \ref{fig:STO} shows the DOS and the TE properties.
 The values of the Seebeck coefficient match the experimentally
 observed ones~\cite{heavydopping1,heavydopping2,heavydopping3} 
 by several authors at the optimal carrier concentration ($10^{21}$ cm$^{-3}$).
 Since the basic electronic structure is very similar,
 this constitutes a strong validation of our calculations 
 for the double perovskites presented so far.

 If we compare the results for double perovskites with the ones with STO, 
 we can observe that the values for the 
 Seebeck coefficient are, in some cases like LaSrTiIrO$_6$ 
 [Fig.~\ref{fig:test}(a)], larger that those for STO.
 The electrical conductivity for these double perovskites 
 is of the same order of magnitude or even higher than that of STO 
 [see, for example,
 LaSrTiIrO$_6$ on KTO in Fig.~\ref{fig:test}(a) 
 or Sr$_2$TaIrO$_6$ in Fig.~\ref{fig:test}(c)],  
 assuming that the scattering time remains similar for both systems,
 which seems a reasonable premise.

 But what makes them ultimately very interesting materials is their very 
 low thermal conductivity compared with that of STO.
 At the optimal doping regime for these compounds, 
 thermal conductivities below 1 W $\times$ m$^{-1}$ $\times$ K $^{-1}$ have
 been reported~\cite{thermal1} for double perovskites,
 while in STO typical values are easily above
 3 W $\times $ m$^{-1}$ $\times$ K $^{-1}$. 
 This can provide a thermoelectric figure of merit up to 3 times 
 larger that that of STO,
 making these compounds strong candidates 
 for being good TE materials.

\section{Summary}

 We have studied various nonmagnetic oxides in a double-perovskite structure,
 with a $d^0$/$d^6$ ionic configuration combining transition metals 
 from different series.
 These compounds show a very large Seebeck coefficient,
 comparable with the largest values calculated for the best TE oxides 
 at reasonable electron-doping levels.
 The driving force for the large $S$ is the existence of a peak in the DOS
 coming from the empty $t_{2g}$ bands of one of the cations at the B-site.
 Our simulations show that the electrical conductivity can be improved by
 the introduction of broad $e_g$ bands from $4d/5d$ cations. 
 Even more, the band structure can be tuned applying epitaxial strain,
 so the power factor can be enhanced by 	up to a 30\% with respect to the 
 relaxed cases within the cubic symmetry in some compounds,
 achieving values even higher than other similar oxides like SrTiO$_3$.
 This, combined with experimental evidences\cite{thermal1} showing that
 several double perovskites have lower ($\sim3$ times) thermal
 conductivity than SrTiO$_3$\cite{heavydopping1,heavydopping2,heavydopping3} at
 the optimal carrier concentrations makes these materials potential
 candidates to become (when optimally doped) good thermoelectrics. 

 In a more general context, we have shown through first-principles
 calculations the effectiveness of the combination of a highly
 degenerate band structure around the Fermi level with a more spread $e_g$
 band which provides good electrical conductivity together with a narrower and
 sharp $t_{2g}$ band to optimize the Seebeck coefficient. 
 The crucial point is to have the maximum degeneracy possible at the 
 bottom of the conduction band (if electron
 doping is considered, as we have done here) but with a lower lying $e_g$
 band providing the power factor enhancement. 
 Our calculations show that this configuration optimizes
 the TE power factor in this kind of systems, 
 and whenever it does not happen as a natural
 configuration, it can be achieved effectively by means of biaxial strain 
 and/or volume reduction,
 if the starting band structure is sufficiently close to the ``ideal'' one.
 
 We also want to stress that in order to make predictions about any specific compound which is proposed here,
 one should perform a calculation including a full relaxation of the atomic structure, which could introduce
 deviations from the ideal configuration (such as distorted or rotated octahedra) that would drive to
 slightly different results coming from a modified electronic structure. Nevertheless, one could find a compound in which a full relaxation of the structure locates
 the $e_g$ and $t_{2g}$ bands in the correct place with respect to each other. Then strain, volume reduction,
 or some of the other methods discussed throughout the text can be used to precisely tune the band structure and reach the optimal TE efficiency.

\acknowledgments

 PVA and VP thank the Xunta de Galicia for financial support through project 
 EM 2013/037. 
 VP, PGF and JJ acknowledge financial support from the Spanish Ministery of
 Economy and Competitiveness through the MINECO Grants
 No. MAT2013-44673-R (VP), and No. FIS2012-37549-C05-04 (PGF and JJ).
 VP and PGF also acknowledge funding from the Ram\'on y Cajal Fellowship
 RYC-2011-09024 and RYC-2013-12515, respectively.

\end{document}